\documentclass[aps,prl,twocolumn,superscriptaddress,groupedaddress]{revtex4}  
\usepackage{graphicx}  
\usepackage{dcolumn}   
\usepackage{bm}        
\usepackage{amssymb}
\usepackage{amsmath} 
\usepackage{subfig}
\usepackage{float} 
\usepackage{lipsum} 

\begin{document}



\title{Oscillating Entropy and Spin Precession in the Ensemble of Qubits Interacting with Thermal Systems}
\author{Xiaoyu He}
\affiliation{Department of Applied Physics, NYU
	Tandon School of Engineering, Brooklyn, NY 11201, USA}
\author{Zain H.Saleem}
\affiliation{Theoretical Research Institute of Pakistan Academy of Sciences,
	Islamabad 44000, Pakistan}
\author{Vladimir I. Tsifrinovich}
\affiliation{Department of Applied Physics, NYU
	Tandon School of Engineering, Brooklyn, NY 11201, USA}
     
\date{\today}

\begin{abstract}
We present a simple model which allows us to explain the physical nature of the oscillating entropy. We consider an ensemble of qubits interacting with thermal two-level systems. The entropy of the qubits oscillates between zero and the value of entropy of the thermal systems. We show that the oscillations of the entropy can be clearly explained by the precession of the real or effective spins of the qubits. 
\end{abstract}

\maketitle

 Interaction between the qubits and their environment is vital for quantum information processing \cite{1}. One of the most important parameters describing an ensemble of qubits interacting with the thermal environment is the von Neumann entropy. In recent years the entropy of the qubits has attracted a lot of attention, especially, in connection with the qubit entanglement \cite{2,3,4,5}. It was found that the qubit entropy can oscillate with time when the two qubits, initially in a superpositional state, interact with a single oscillator \cite{6}. The entropy also oscillates when the two qubits, initially in the entangled state, interact with the two independent photon baths in isolated cavities \cite{7}. In both cases the whole system (qubits plus a system interacting with the qubits) is in the pure quantum state.
 
While the phenomenon of the entropy oscillations have been described in \cite{6,7} the physical nature of these oscillations remained obscure. In this work we suggest a simple model which clearly demonstrates the nature of the entropy oscillations. In our model a single qubit spin (Q-spin) interacts with a single thermal spin (T-spin). We will show that the the qubit entropy oscillates with time between zero and the value of the entropy of the thermal system. The entropy of the thermal system does not change for the Ising interaction and remains, approximately, constant for the Heisenberg interaction. Thus, the entropy oscillations cannot be associated with the flow of entropy between the qubit and the thermal system. We will show that the entropy oscillations can be clearly explained by precession of the Q-spins. 

 First, we consider the Ising interaction between the Q-spins and the T-spins. The Hamiltonian for the Q-T-spins can be written in terms of the Pauli operators
\begin{equation}
H=-E_1\sigma_{z1}-E_2 \sigma_{z2} -J\sigma_{z1}\sigma_{z2}. 
\label{eq1}
\end{equation}
Here $E_1$ and $E_2$ are the Zeeman energies of the Q- and T-spins, respectively, and $J$ is the interaction constant. We assume that initially the qubits are placed into the uniform superposition of stationary states so that the Q-spins point in the positive x-direction. The corresponding density matrix is,
\begin{equation}
\rho_1(0) = \frac{1}{2} \left(
\begin{array}{cc}
1 & 1 \\
1 & 1 \\
\end{array}
\right).
\end{equation}

The initial density matrix of the T-spins at temperature T is given by the expression: $\rho_2(0)= \text{diag}(f_{00},f_{11}) $, where $f_{00}= Z^{-1} e^{-E_2 T}$, $f_{11}= Z^{-1} e^{E_2/ T}$,  and $Z$ is the partition function $Z= e^{-E_2 T}+ e^{E_2 T}$. (Here and below we put $k_B = 1$ and $\hslash = 1$.) The $4\times4$ density matrix of the two systems is the tensor product: $\rho(0)= \rho_1(0) \otimes \rho_2(0) $.
Solving the von Neumann equation, 
\begin{equation}
i \dot{\rho} (t)= [H,\rho(t)],
\end{equation}
 we obtain the time dependent density matrix with the non-zero components,  
 \begin{eqnarray}
 \rho_{00}&=&\rho_{22} =\frac{1}{2} f_{00}, \;\;\; \rho_{02}=\frac{1}{2} f_{00}e^{2i(E_1+J)t}, \nonumber \\\rho_{11}&=&\rho_{33}= \frac{1}{2} f_{11}, \;\;\; \rho_{13}= \frac{1}{2}f_{11}e^{2i(E_1-J)t}.
 \end{eqnarray}Tracing the density matrix over the thermal system we find the reduced density matrix of the quibit's ensemble:
\begin{equation}
\rho_1(t)= Tr_2 \rho(t) = \frac{1}{2} \left(
\begin{array}{cc}
1 & a \\
a^* & 1 \\
\end{array}
\right),
\end{equation}
\begin{equation}
a= f_{00} e^{2 i (E_1+J)t}+ f_{11} e^{2 i (E_1-J)t} \nonumber.
\end{equation}
The reduced density matrix of the thermal ensemble obtained by tracing over the qubits does not change: $\rho_2(t)=\rho_2(0)$.

 The von Neumann entropy of the qubits with the Ising interaction is given by the expression:
\begin{eqnarray}
S_1(t)&=& -Tr \{\rho_1(t) \ln \left( \rho_1(t)\right)\}\\
&=&\frac{1}{2} X \ln \left(\frac{1-X}{1+X}\right)
-\frac{1}{2} \ln
\left(f_{00} f_{11} \sin ^2(2 J t)\right), \nonumber
\end{eqnarray}
where, 
\begin{equation}
X=(1-4f_{00} f_{11} \sin^2 (2 J t))^{1/2}.
\end{equation}
\vspace{5mm}

\begin{figure}[H]
\includegraphics[scale=0.55]{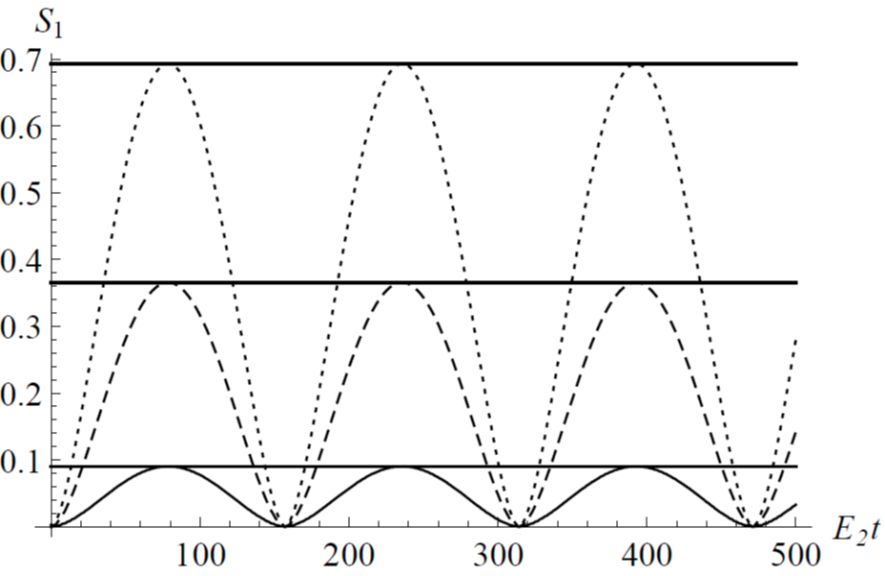}
\caption{\label{fig1}Qubit entropy $S_1$ as a function of the dimensionless time $E_2 t$ for the following values of parameters:  $E_2=1$, $E_1=10^{-4}$ and $J=10^{-2}$. The solid line corresponds to the temperature $T=0.5$ , the dashed line $T=1$, and the dotted line $T=\infty$. The horizontal lines on the graph show the values of the entropy of the thermal system.}
\end{figure}

We will assume the following relation between the parameters of the Hamiltonian: $E_1\ll J \ll E_2$. In this case the entropy (6) oscillates between zero and the value of the entropy of the thermal system. (See Fig. \ref{fig1}.) On the other hand, the entropy of the thermal system $S_2$ does not change with time:
\begin{eqnarray}
S_{2}&=&-Tr \{\rho_2 \ln \rho_2\}\nonumber \\
&=&-( f_{00} \ln {f_{00}} + f_{11}\ln {f_{11}}).
\end{eqnarray}

\begin{figure}[H]
\includegraphics[scale=0.61]{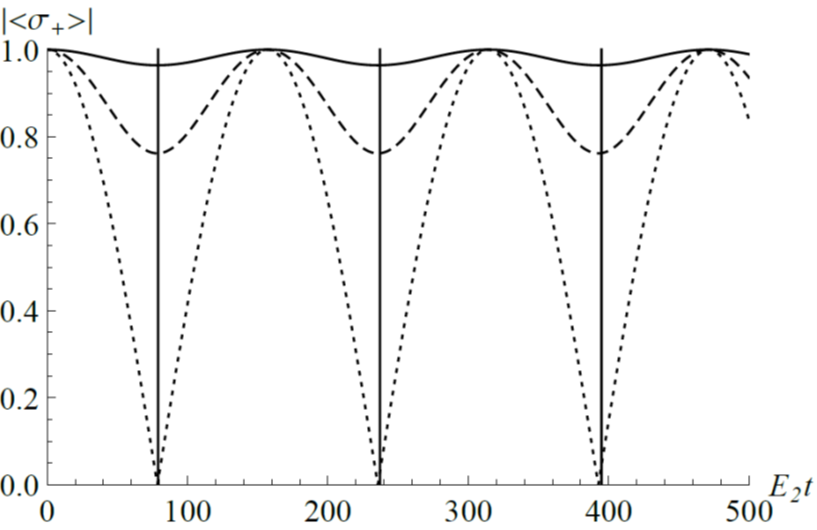}
\caption{\label{precI1}The  absolute value (amplitude)  of  the  Q-spin        precession $|<\sigma_+(t)>|$ as a function of the dimensionless time for the same values of parameters as in Fig. 1. The vertical lines show the positions of the maxima of entropy.}
\end{figure}

In order to explain the entropy oscillations we will consider the precession of the Q-spins. The transversal component of the Q-spins computed with the reduced density matrix (5) is exactly the same as that computed with the full density matrix (4):
\begin{equation}
\begin{split}
<\sigma_+>&= <\sigma_x + i \sigma_y>= Tr\{\sigma_+\rho_1(t)\}\\
&=f_{00}e^{-2i(E_1+J)t}+f_{11}e^{-2i(E_1-J)t}.\\
\end{split}
\label{eq2}
\end{equation}
 The graph of the precession amplitude $|<\sigma_+(t)>|$ is shown in Fig. \ref{precI1} for the same values of parameters as in Fig. \ref{fig1}. One can see that the maxima of the precession amplitude correspond to the zero entropy and the minima correspond to the maximum entropy.

\begin{figure}[H]
	\includegraphics[scale=.55]{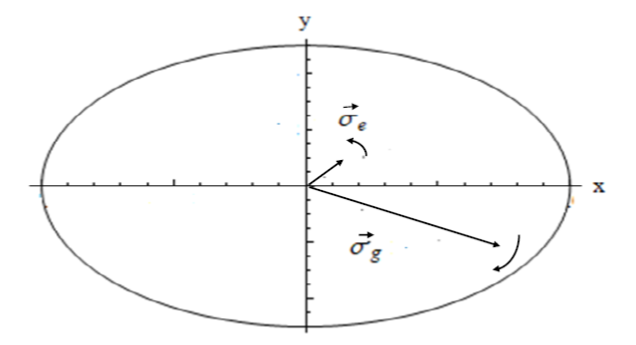}
	\caption{\label{polarI}Precession of the Q-spins. The long arrow $\vec{\sigma_e}$ shows the sum of the Q-spins interacting with the T-spins in the ground state, the short arrow $\vec{\sigma_e}$ is the same with the T-spins in the excited state.}
\end{figure}

The physical reason of this correspondence is the following. Assume, for simplicity, that $E_1=0$. Then, as one can see from Eq. (5), the precession frequency of all Q-spins in the ensemble is the same and equals $2J$. The precession is clockwise when the T-spin, interacting with the Q-spin, is in its ground state and counterclockwise in the opposite case. At zero temperature all the Q-spins precess clockwise, i.e. the precession is circular polarized. At any instant of time all the Q-spins point in the same direction (perfect order). Correspondingly the entropy of the Q-spins equals zero. At a finite temperature, a Q-spin precesses clockwise with the probability $f_{00}$ and  counterclockwise with the probability $f_{11}$ i.e. the precession is elliptically polarized. (See Fig. \ref{polarI}.) 
At instants $2Jt=0, \;\pi,\; 2\pi,...$ all the Q-spins point in the same direction along the x-axis, and the qubit entropy equals zero. At instants $2Jt=\pi/2,\;3\pi/2,\;5\pi/2,...$ the fraction $f_{00}$ of the Q-spins points in the positive or negative y-direction while the fraction $f_{11}$ of the Q-spins points in the opposite direction. At these instants the Q-spins are in the state of the maximum disorder, and their entropy will be the same as the entropy of the thermal system: $S_1=S_2= -( f_{00} \ln{f_{00}} + f_{11}\ln {f_{11}})$. In the extreme case of the infinite temperature the precession of the Q-spins becomes linearly polarized, and the maximum entropy of the Q-spins is $S_1=S_2=\ln 2$.  

One could expect that the described situation is unique and caused by the strict conservation of the z-components of the Q- and T-spins for the Ising interaction. Below we will show that the same phenomenon remains if we replace the Ising interaction $-J\sigma_{z1}\sigma_{z2}$  with the Heisenberg one $- J \; \vec{\sigma_1}\cdot \vec{\sigma_2}$. For the Heisenberg interaction the z-components of Q- and T-spins do not conserve. A qubit can exchange energy with the thermal system.

The components of the density matrix $\rho_{ij}$ for the whole Q-T-system with the Heisenberg interaction are given in Eq.(15) in the Appendix. The components of the reduced density matrix for the thermal system, which we denote $b_{ij}$, are given in Eq. (16) in the Appendix. One can see that the components $b_{ij}$ of the reduced density matrix for the T-spins oscillate with time. However for our relation between the parameters $E_1\ll J\ll E_2$, the amplitude of oscillations of the diagonal components $b_{ii}$ is of the order of $(J/E_2)^2 \ll 1$.
 
 The entropy of the thermal system $S_2(t)$ is given by the expression,
\begin{equation}
S_2(t)= \frac{1}{2}\ln\;4 -\frac{1}{2}\ln(1-{X_2}^2) +X_2 \ln\left(\frac{1-X_2}{1+ X_2}\right),\\
\end{equation}
where $W^2= E_{12}^2 + 4 J^2$, $E_{12}=(E_1-E_2)$ and 
\begin{eqnarray}
 {X_2}&=& \Big(\frac{(f_{00} -f_{11})^2 ( W^2- 4J^2 \sin^2( tW))^2}{W^4} \\
&+&\frac{4 J^2 (1- 4 f_{00}f_{11}\cos^2(2Jt))\sin^2(2t W) }{W^2}\Big)^{1/2}. \nonumber
\end{eqnarray}

The components of the reduced density matrix for the qubits, which we denote $a_{ij}$, are given in Eq. (17) in the Appendix. One can see that the diagonal components $a_{ii}$ oscillate with the amplitude of the order of $(J/E_2)^2 \ll 1$. 

The qubit entropy in the Heisenberg interaction $S_1$ is described by the expression:
\begin{eqnarray}
S_1(t)&=&  - Tr \{ \rho_1(t) \ln \left(\rho_1 (t)\right) \} \\
&=&\frac{1}{2} X_1 \ln \left(\frac{1-X_1}{1+X_1}\right)-\frac{1}{2} \ln (Y_1).\nonumber
\end{eqnarray}
where 

\begin{eqnarray}
{X_1}&=&\Big( 1-4Y_1+ 4 \frac{J^4 (f_{00}-f_{11})^2 \sin ^4(t W)}{W^4}\Big)^{1/2},\\
Y_1&=& f_{00} f_{11} \sin ^2(2 J t)-\frac{4 J^4 (f_{00}-f_{11})^2 \sin ^4(t W)}{W^4} \nonumber \\ &+& \frac{J^2 \sin ^2(t W) \left(1-4 f_{00} f_{11} \sin ^2(2 J t)\right)}{W^2}. \nonumber
\end{eqnarray}

The transversal component of the Q-spins for the Heisenberg interaction is found to be,
\begin{eqnarray}
&&  <\sigma_+> = Tr \{ \sigma_+ \rho_1(t)\}= (E_{12}+W)
e^{ -it (E_{1}+E_{2})}  \nonumber \\
&\times & \big(\frac{f_{00} e^{ -it (2 J+W)}+ f_{11} e^{-it (-2 J+W)}}{2W} \nonumber
\\
&-&\frac{f_{00} e^{-it (2 J-W)}+ f_{11} e^{ -it (-2 J-W)}}{2W}\big) 
\end{eqnarray}

For our relations between the parameters $E_1\ll J \ll E_2$ we have $X_1\approx X$, $X_2 \approx f_{00}-f_{11}$ and the expressions (10), (12), and (14), obtained for the Heisenberg interaction, transform to the corresponding formulas (8), (6), and (9), obtained for the Ising interaction. Thus for the Heisenberg interaction the entropy oscillations can be explained by the Q-spin precession in the same way as for the Ising interaction.

In conclusion, we have presented a simple model where the oscillations of the qubit entropy can be clearly explained by the precession of the real or effective spins of the qubits. The entropy is zero when all the Q-spins point in the same direction, and it is maximum when the Q-spins, interacting with the T-spins in the ground and excited states, point in the opposite directions.

In the case of the Ising interaction between the Q- and T-spins the state of the thermal system does not change with time. The Heisenberg interaction causes the energy exchange between the Q- and the T-spins. However, for the range of parameters considered in our work, the oscillations of the qubit entropy and precession amplitude of the Q-spins are almost the same for the both interactions. 

The presented model can describe, for example, evolution of electron and nuclear spins of impurity atoms coupled by the hyperfine interaction, during the time interval small compared to the time of electron spin relaxation.
\vspace{3mm}

The authors thank Prof L.M. Folan for useful discussions.

\section{Appendix}
For the Heisenberg interaction the components $\rho_{ij}$ of the density matrix for the whole Q-T-system are given by the expressions,

\begin{widetext}

		\begin{eqnarray}
		\rho_{01}&=&- i  J  f_{00} \frac{ e^{i t (E_1+E_2+2 J)} \sin \left(t W \right)}{W}, \;\;
		\rho_{02}=\frac{ f_{00} e^{i t (E_1+E_2+2 J)} \left(W \cos \left(t W \right)+i
			E_{12} \sin \left(t W\right)\right)}{2 W},  \nonumber\\
		\rho_{11}&=&\frac{ f_{11}  \left( 2 J^2 \cos \left(2 t W\right)+E_{12}^2+2 J^2\right)+ 4 f_{00} J^2
			\sin ^2\left(t W\right)}{2W^2}, \;\;\;\;\;\;\rho_{00} = \frac{1}{2} f_{00}, \nonumber \\ \rho_{22} &=& \frac{ f_{00}  \left(2 J^2 \cos \left(2 t W\right)+E_{12}^2+2 J^2\right)+ 4 f_{11} J^2
			\sin ^2\left(t W\right)}{2W^2}, \;\;\;\;\;\;\rho_{33}=\frac{1}{2}  f_{11}, \nonumber \\
		\rho_{12} &=& \frac{1}{2}\frac{J \left( f_{00} - f_{11} \right) \left(i W \sin \left(2 t
			W\right)+E_{12} \cos \left(2 t W \right)-E_{12}\right)}{W^2},  \;\;\;
		\rho_{23}= \frac{ i  f_{11} J  e^{i t (E_1+E_2-2 J)} \sin \left(t W\right)}{W}, \nonumber \\ 
		\rho_{13}&=&\frac{f_{11} e^{i t (E_1+E_2-2 J)} (W \cos (t W)+i E_{12} \sin (t W))}{2 W}, \;\;\;\;\;\;\;\;\;\;\;\;\;\;\; \rho_{03}=  0.
		\end{eqnarray}
\end{widetext}
where $W^2= E_{12}^2 + 4 J^2$. The components of the reduced density matrix for the T-spins are given by

	\begin{eqnarray}
	b_{00}&=& \frac{1}{2} f_{00} + \frac{E_{12}^2 f_{00} + 2 J^2 + 2 (f_{00}-f_{11})J^2 \cos{(2t W)}}{2W^2},  \nonumber\\
	b_{11}&=& \frac{1}{2} f_{11} + \frac{E_{12}^2 f_{11} + 2 J^2 + 2 (f_{11}-f_{00})J^2 \cos{(2t W)}}{2W^2}, \nonumber \\
	b_{01}&=& \frac{ iJe^{-i(E_1+E_2+ 2J)t}(f_{00}-e^{4iJt}f_{11})\sin(t W)}{2W}.
	\end{eqnarray}
\normalsize{The components of the reduced density matrix for the Q-spins are}
 
	\begin{eqnarray}
	a_{00}&=& \frac{1}{2}f_{00} + \frac{E_{12}^2 f_{11} + 2 J^2 + 2 (f_{11}-f_{00})J^2 \cos{(2t W)}}{2W^2}, \nonumber \\
	a_{11}&=& \frac{1}{2} f_{11} + \frac{E_{12}^2 f_{00} + 2 J^2 + 2 (f_{00}-f_{11})J^2 \cos{(2t W)}}{2W^2}, \nonumber \\
	a_{01}&=& e^{i t (E_1+E_2+2 J)} \left(f_{00}+f_{11} e^{-4 i J t}\right) \times \nonumber \\ &&	\frac{ (W \cos (t W)+i E_{12} \sin (t
		W))}{2W}.
	\end{eqnarray}

\input    

\end{document}